\input harvmac
\input epsf

\newcount\figno
\figno=0
\def\fig#1#2#3{
\par\begingroup\parindent=0pt\leftskip=1cm\rightskip=1cm\parindent=0pt
\baselineskip=11pt
\global\advance\figno by 1
\midinsert
\epsfxsize=#3
\centerline{\epsfbox{#2}}
\vskip 12pt
{\bf Fig. \the\figno:} #1\par
\endinsert\endgroup\par
}
\def\figlabel#1{\xdef#1{\the\figno}}
\def\encadremath#1{\vbox{\hrule\hbox{\vrule\kern8pt\vbox{\kern8pt
\hbox{$\displaystyle #1$}\kern8pt}
\kern8pt\vrule}\hrule}}

\overfullrule=0pt

\noblackbox
\parskip=1.5mm

\def\Title#1#2{\rightline{#1}\ifx\answ\bigans\nopagenumbers\pageno0
\vskip0.5in
\else\pageno1\vskip.5in\fi \centerline{\titlefont #2}\vskip .3in}


\noblackbox
\parskip=1.5mm

  
\def\npb#1#2#3{{\it Nucl. Phys.} {\bf B#1} (#2) #3 }
\def\plb#1#2#3{{\it Phys. Lett.} {\bf B#1} (#2) #3 }
\def\prd#1#2#3{{\it Phys. Rev. } {\bf D#1} (#2) #3 }

\def\mpla#1#2#3{{\it Mod. Phys. Lett.} {\bf A#1} (#2) #3 }

\def\ijmpd#1#2#3{{\it Int. J. Mod. Phys.} {\bf D#1} (#2) #3 }

\def\bb#1{{\tt hep-th/#1}}
\def\grqc#1{{\tt gr-qc/#1}}

\def\app#1#2#3{{\it Astropart. Phys. } {\bf #1} (#2) #3 }
\def\cqg#1#2#3{{\it Class. Quant. Grav. } {\bf #1} (#2) #3 }

\def\aop#1#2#3{{\it Annals Phys.} {\bf #1} (#2) #3 }
\def\jhep#1#2#3{{\it JHEP } {\bf #1} (#2) #3 }
\def\prep#1#2#3{{\it Phys. Rept.} {\bf #1} (#2) #3 }


           \def\CO{{\cal O}}

 \def\CR{{\cal R}} \def\CD{{\cal D}}

\def\stackrel#1#2{\mathrel{\mathop{#2}\limits^{#1}}}


\def\dj{\hbox{d\kern-0.347em \vrule width 0.3em height 1.252ex depth
-1.21ex \kern 0.051em}}

\def\half{{1\over 2}\,}
\def\d{{\rm d}}


\lref\rwittkk{E. Witten, \npb{186}{1981}{412}; {\it Fermion quantum numbers
in Kaluza-Klein theory}, Proceedings of the Shelter Island II conference,
M.I.T. Press, 1985.}
\lref\ralgo{M. Carmeli, C. Charach and A. Feinstein, \aop{150}{1983}{392.}}
\lref\rgh{G.W. Gibbons and P.K. Townsend, \npb{282}{1987}{610\semi}
G.W. Gibbons, G.T. Horowitz and P.K. Townsend, \cqg{12}{1995}{297} 
(\bb{9410073}).}
\lref\rlw{F. Larsen and F. Wilczek, \prd{55}{1997}{4591} (\bb{9610252}).}  
\lref\rbfm{T. Banks, W. Fischler and L. Motl, \jhep{01}{1999}{019} 
(\bb{9811194}).}
\lref\rrabi{O. Aharony, \npb{476}{1996}{470} (\bb{9604103})\semi
S. Elitzur, A. Giveon, D. Kutasov and E. Rabinovici, \npb{509}{1998}{122}
(\bb{9707217}).}
\lref\rms{J. Maharana and J.H. Schwarz, \npb{390}{1993}{3} (\bb{9207016}).}
\lref\rours{A. Feinstein, R. Lazkoz and M.A. V\'azquez-Mozo, 
\prd{56}{1997}{5166} (\bb{9704173}).}
\lref\racf{T. Appelquist, A. Chodos and P.G.O. Freund, {\it Modern
Kaluza-Klein Theories}, Addison Wesley 1987.}
\lref\rmt{H. L\"u, S. Mukherji, C.N. Pope and K.-W. Xu, \prd{55}{1997}{7926}
(\bb{9610107})\semi
A. Lukas, B.A. Ovrut and D. Waldram, \npb{495}{1997}{365} 
(\bb{9610238})\semi
A. Lukas and B. Ovrut, \plb{437}{1998}{291} (\bb{9709030})\semi
N. Kaloper, I. Kogan and K.A. Olive, \prd{57}{1998}{7340} (\bb{9711027})\semi
K. Benakli, \ijmpd{8}{1999}{153} (\bb{9804096})\semi
A. Lukas, B.A. Ovrut and D. Waldram, {\it Cosmolgical solutions of 
Horava-Witten theory}, \bb{9806022}\semi
H.S. Reall, \prd{59}{1999}{103506} (\bb{9809195})\semi
M. Maggiore and A. Riotto, {\it D-branes and cosmology}, \bb{9811089}.}
\lref\rkkpr{J.M. Overduin and P.S. Wesson, \prep{283}{1997}{303} (\bb{9805018}).}
\lref\rbbang{E. Alvarez, \prd{31}{1985}{418\semi}
R. Brandenberger and C. Vafa, \npb{316}{1989}{191\semi}
K.A. Meissner and G. Veneziano, \plb{267}{1991}{33;} \mpla{6}{1991}{3397} 
(\bb{9110004})\semi
M. Gasperini, J. Maharana and G. Veneziano, \plb{296}{1992}{51} (\bb{9209052})\semi
M. Gasperini and G. Veneziano, \app{1}{1997}{317} (\bb{9211021})\semi
N. Kaloper R. Madden and K.A. Olive, \npb{452}{1995}{677} (\bb{9506027})\semi 
M. Gasperini, M. Maggiore and G. Veneziano, \npb{949}{1997}{315} (\bb{9611039})\semi
R. Brandenberger, R. Easther and J. Maia, \jhep{08}{1998}{007} (\grqc{9806111})\semi
D.A. Easson and R. Brandenberger, {\it Nonsingular dilaton cosmology in the string 
frame}, (\bb{9905175}).}
\lref\rrm{M. Mueller, \npb{337}{1990}{37\semi}
G. Veneziano, \plb{265}{1991}{287\semi}
A.A. Tseytlin and C. Vafa, \npb{372}{1992}{443} (\bb{9109048})\semi
A.A. Tseytlin, \plb{334}{1994}{315} (\bb{9404191})}
\lref\rbf{K. Behrndt and S. F\"orste, \plb{320}{1994}{253} (\bb{93081331});
\npb{430}{1994}{441} (\bb{9403179})}
\lref\rdc{A. Chodos and S. Detweiler, \prd{21}{1980}{2167.}}
\lref\rbm{M. Gleiser and J.A. Stein-Schabes, \prd{34}{1986}{1739.}}
\lref\rlp{H. L\"u and C.N. Pope, \npb{465}{1996}{127} (\bb{9512012}).}
\lref\rop{N.A. Obers and B. Pioline, {\it U-duality and M-theory},
\bb{9809039}, to appear in Physics Reports.}
\lref\rkkbm{P.G.O. Freund, \npb{209}{1982}{146\semi}
D. Sahdev, \plb{137}{1984}{155\semi}
S. Randjbar-Daemi, A. Salam and J. Straathdee, \plb{135}{1984}{388}}
\lref\rdem{J. Demaret and J.-L. Hanquin, \prd{31}{1985}{258\semi}
J. Demaret, J.-L. Hanquin, M. Henneaux and P. Spindel, 
\npb{252}{1985}{538.}}
\lref\rtm{E.J. Copelan, A. Lahiri and D. Wands, \prd{50}{1994}{4868} 
(\bb{9406216})\semi
N.A. Batakis and A.A. Kehagias, \npb{449}{1995}{248} 
(\bb{9502007})\semi
N.A. Batakis,\plb{353}{1995}{450} (\bb{9503142}); \plb{353}{1995}{39} 
(\bb{9504057})\semi
J.D. Barrow and K.E. Kunze, \prd{55}{1997}{623} (\bb{9608045}); 
\prd{56}{1997}{741} (\bb{9701085})\semi 
J.D. Barrow and M.P. Dabrowski, \prd{55}{1997}{630} (\bb{9608136}).}
\lref\raqmw{D. Clancy, A. Feinstein, J. Lidsey and R. Tavakol, {\it
Inhomogeneous Einstein-Rosen string cosmologies}, (\grqc{9901062}).}


\line{\hfill EHU-FT-9910}
\line{\hfill ITFA-99-11}  
\line{\hfill SPIN-1999/16}
\line{\hfill {\tt hep-th/9906006}}

\vskip -0.8cm

\Title{\vbox{\baselineskip 12pt\hbox{}
 }}
{\vbox{\centerline{M-theory resolution of four-dimensional}
{\centerline{    }}
{\centerline{cosmological singularities via U-duality}}
}}

\vskip 0.3cm

\centerline{$\quad$ {A. Feinstein$^{\, \rm a,}$\foot{wtpfexxa@lg.ehu.es} and 
M.A. V\'azquez-Mozo$^{\, \rm b,c,}$\foot{vazquez@wins.uva.nl, 
M.Vazquez-Mozo@phys.uu.nl}}}

\medskip

\centerline{{\sl $^{\rm a\,}$Dpto. de F\'{\i}sica Te\'orica}}
\centerline{{\sl Universidad del Pa\'{\i}s Vasco}}
\centerline{{\sl Apdo. 644, E-48080 Bilbao, Spain}}

\vskip 0.3cm

\centerline{{\sl $^{\rm b\,}$Instituut voor Theoretische Fysica}}
\centerline{{\sl Universiteit van Amsterdam, Valckenierstraat 65}} 
\centerline{{\sl 1018 XE Amsterdam, The Netherlands}}

\vskip 0.3cm

\centerline{{\sl $^{\rm c\,}$Spinoza Instituut, Universiteit Utrecht}}
\centerline{{\sl Leuvenlaan 4, 3584 CE Utrecht, The Netherlands}}

\vskip 0.5cm

\noindent
We consider cosmological solutions of string and M-theory compactified to 
four dimensions by giving a general prescription to construct 
four-dimensional modular cosmologies with two commuting Killing vectors
from vacuum solutions. By lifting these solutions to higher dimensions 
we analyze the existence of cosmological singularities and find that, 
in the case of non-closed Friedmann-Robertson-Walker universes, 
curvature singularities are removed from the higher-dimensional 
model when only one of the extra dimensions is time-varying. By 
studying the moduli space of compactifications of M-theory resulting 
in homogeneous cosmologies in four dimensions we show that U-duality 
transformations map singular cosmologies into non-singular ones.


\Date{6/99}


\newsec{Introduction and motivation}

It is well known that string physics has changed the way we now look
at Cosmology. In particular, string theory leads to consider 
multidimensional cosmological scenarios in a natural way, since
superstrings can only be consistently quantized in ten dimensions.
M-theory has confirmed this trend extending cosmology to the 
realm of the eleventh dimension (see for example \refs\rmt). 

Higher dimensional cosmology is not new. Before the
advent of string theory the Kaluza-Klein paradigm had already put forward
the idea that four-dimensional space-time should not be something
taken for granted in cosmology (a collection of the most relevant articles on
the subject can be found in \refs\racf; see also \refs\rkkpr). In its 
original version, the Kaluza-Klein program aimed to describe all 
four-dimensional matter fields as 
purely gravitational (or supergravitational) degrees of freedom in 
$4+N$ dimensions. Although this idea met serious obstacles when trying 
to account for the existence of chiral matter in four dimensions 
\refs\rwittkk, it has been partially incorporated in string theory
where the presence of ten-dimensional chiral matter fermions solves 
the problem.

When looking at the four-dimensional physics, the extra internal
dimensions leave their blueprints in the form of a plethora of
massless scalar and vector fields. On general grounds, the presence of these 
massless scalars poses a serious problem in trying to extract realistic 
cosmological models from string theory. The situation is complicated 
by the fact that most of these fields represent flat directions of the 
superpotential that are not lifted by quantum corrections as long as enough
supersymmetry is preserved. In many ocassions, however, the study is 
restricted to the simplest cosmological models by looking just at the effects
of the dilaton and tensor fields, for example, and ignoring the 
remaining moduli fields altogether\foot{From the M-theory perspective, 
on the other hand, the dilaton itself is just a moduli associated with 
the compactification of the eleventh dimension. The reason to separate 
it from other fields is that in the weakly coupled string limit of M-theory 
its compactification scale $g_{s}\ell_{s}$ is much smaller than the 
string scale $\ell_{s}$.} \refs\rtm\refs\rours.

The ultimate motivation of string/M-theory cosmology is of course to solve
the problem of cosmological singularities, explain the initial conditions 
in cosmology and the dimensionality 
of space-time. Close to $t=0$, quantum gravity 
effects become dominant and hopefully will smear the semiclassical 
singularity, thus opening a new window to the study of the early universe.
Although a full quantum cosmology description of the early universe is 
still missing, the semiclassical analysis supported by the use of 
stringy symmetries has been useful in getting a flavor of the physics 
close to the Big-Bang singularity \refs\rbbang\refs\rbfm.
On the other hand, from a Kaluza-Klein perspective it is 
possible that the existence of the initial singularity
might just be the result of integrating out the physics associated
with higher dimensions \refs\rgh\refs\rlw\ and that it could be
removed already in the semiclassical approximation.

In this paper we will combine these two ideas and investigate the 
effect of extra dimensions on the initial cosmological singularity 
by studying a family of Friedmann-Robertson-Walker (FRW) 
cosmologies coupled to a number of scalar fields which we associate 
with compactification moduli. We find that in the particular case of 
open or flat ($k=-1,0$) universes in which only one of the extra dimensions
has a non-trivial dynamics, the Big-Bang singularity might be just an 
artifact of the Kaluza-Klein reduction which may be removed when going 
to higher dimensions. On the other hand, in the case of spatially closed 
solutions ($k=1$) the Big-Bang or the Big-Crunch
singularity is postponed in the higher-dimensional model, 
again when only one of the extra dimensions is time-varying. It is 
interesting that by switching on more than one dynamical internal dimension 
the regularization of curvature singularities is definitively spoiled 
in all cases. When interpreting, in the open and flat cases, 
this higher dimensional cosmologies as solutions in M-theory we find 
that the U-duality group ${\bf G}_{7}$ acts on the moduli space of 
metrics by relating singular geometries with non-singular ones. 
This seems to indicate that a certain class of cosmological singularities
in M-theory can be physically resolved in terms of an equivalent dual
regular background.

The plan of the article is the following: in the next section we 
will review the properties of moduli fields arising from compactifications
of $(4+N)$-dimensional Einstein gravity on a $N$-dimensional straight torus
and provide a general algorithm to construct Gowdy-type cosmologies
in the presence of moduli fields. We will apply in Sec. 3 this algorithm
to generate modular FRW cosmologies and, after undoing the Kaluza-Klein 
reduction, will study the structure of singularities of the higher dimensional
``parent'' metric. In Sec. 4 the analysis will be focused on M-theory 
compactifications to four dimensions and the action of U-duality on the
moduli space of solutions. Finally, conclusions will be summarized in 
Sec. 5.

\newsec{Cosmologies coupled to scalar fields vs. dimensional reduction}

\subsec{Scalar fields from dimensional reduction}

Scalar fields appear naturally in the old
Kaluza-Klein program or its string/M-theory versions. Here, we will 
consider the dimensional reduction of a $(4+N)$-dimensional metric
on an $N$-dimensional straight torus ${\bf T}^{N}=({\bf S}^{1})^{N}$ 
using the ansatz\foot{For a generic analysis of the structure of the
dimensionally reduced action, see \refs\rlp\refs\rop.}
\eqn\kkan{
\eqalign{
ds^2_{4+N} &= e^{-{2\over \sqrt{3}}\sum_{i=1}^{N}\psi_{i}}\, ds^{2}_{4}
+\sum_{i=1}^{N}e^{{4\over \sqrt{3}}\psi_{i}} (dw^{i})^2.
}
}
The metric functions $\psi_{i}(x^{\mu})$ ($\mu=0,\ldots,3$) should be 
thought of as  the components
of the metric in the internal $N$-dimensional torus. Notice that this
ansatz is invariant under the permutation of the $\psi_{i}$ fields.

{}From the four-dimensional point of view, the fields $\psi_{i}(x^{\mu})$ are 
scalars. Their dynamical equations are obtained by demanding
that the $(4+N)$-dimensional metric \kkan\ is a vacuum solution of the
Einstein equations. Writing the Einstein-Hilbert
action for \kkan\ we find that
\eqn\psis{
S=\int d^{4}x \sqrt{-g^{(4)}} \left[\CR^{(4)}-2\left(\sum_{i=1}^{N}
\partial_{\mu}\psi_{i}\partial^{\mu}\psi_{i}+{2\over 3}\sum_{i<j}^{N}
\partial_{\mu}\psi_{i}\partial^{\mu}\psi_{j}\right)\right],
}
leading to the result  that the breathing modes of the higher dimensional
metric appear as {\it mixed} scalar in the four-dimensional action.
 To eliminate this mixing we may perform a diagonalization in
field space by defining the new fields $\varphi_{i}$ through the relation
\eqn\d{
\psi_{i}={\CD}_{ij}\varphi_{j}
}
where $\CD_{ij}\in GL(N,{\bf R})$ is given by
\eqn\d{
\CD=\left(\matrix{\mu_{1}^{-\half} & \mu_{2}^{-\half} & \mu_{3}^{-\half}
& \ldots & \mu_{N-1}^{-\half} & \mu_{N}^{-\half} \cr
-\mu_{1}^{-\half} & \mu_{2}^{-\half} & \mu_{3}^{-\half} & \ldots &
\mu_{N-1}^{-\half} & \mu_{N}^{-\half} \cr
0 & -2\mu_{2}^{-\half} & \mu_{3}^{-\half} & \ldots &
\mu_{N-1}^{-\half} & \mu_{N}^{-\half} \cr
\vdots & \vdots & \vdots & & \vdots & \vdots \cr
0 & 0 & 0 & \ldots & \mu_{N-1}^{-\half} & \mu_{N}^{-\half} \cr
0 & 0 & 0 & \ldots & -(N-1)\mu_{N-1}^{-\half} & \mu_{N}^{-\half}}
\right)
}
with 
$$
\eqalign{
\mu_{n}&={2\over 3}n(n+1), \hskip1cm n=1,\ldots,N-1 \cr
\mu_{N}&={1\over 3}N(N+2),
}
$$
which {\it is not} an $O(N)$ transformation.
The new scalar fields $\varphi_{i}$ will be propagation 
eigenstates and there will be no classical mixing among those. Their
dynamics will be governed by the action
\eqn\ac{
S=\int d^4 x \sqrt{-g^{(4)}}\left[\CR^{(4)}-
2\sum_{i=1}^{N}\partial_{\mu}\varphi_{i}\,
\partial^{\mu}
\varphi_{i}\right].
}
 
In our analysis we have assumed that {\it all} scalar fields
have a geometric origin as moduli of dimensional reduction. This point
of view is very much appropriate for M-theory where the dilaton is
on the same footing with all other scalar as compactification
moduli. Nonetheless, in those regimes of M-theory that can be described in 
terms of a weakly coupled string theory the dilaton field plays a privileged
role as the field whose vacuum expectation value determines the string
coupling constant. In this case, dimensional reduction from ten dimensions
will produce again a number of scalar fields in the lower-dimensional
theory. The dynamics of those moduli can be extracted again from the
Kaluza-Klein ansatz 
\eqn\aten{
ds^2_{10}=ds^2_{4}+\sum_{i=1}^{6}e^{2\sqrt{2}\sigma_{i}}(dw^{i})^{2}
}
where now the ten-dimensional metric is no longer a vacuum solution of 
Einstein equation but rather a solution of dilaton gravity instead. The resulting
four-dimensional action in string frame is \refs\rms
\eqn\st{
S_{\rm string}=\int d^{4}x \sqrt{-g^{(4)}}\,e^{-2\phi}\,
\left[\CR^{(4)}+4\partial_{\mu}
\phi\,\partial^{\mu}\phi-2\sum_{i=1}^{6}\partial_{\mu}\sigma_{i}\,
\partial^{\mu}\sigma_{i}\right]
}
where the four-dimensional dilaton $\phi$ is defined in terms of the 
ten-dimensional one as
\eqn\tdd{
\phi=\phi_{(10)}-{1\over\sqrt{2}}\sum_{i=1}^{6}\sigma_{i}.
}
We see from \st\ that the kinetic term for the fields $\sigma_{i}$ is
diagonalized in the four dimensional action, although it is conformally 
coupled to the dilaton field. This conformal coupling can be removed as 
usual by a conformal 
transformation of the metric. It is however when we re-express the 
four-dimensional dilaton in terms of the ten-dimensional one via \tdd\
that the mixing between the different $\sigma_{i}$ appears. In order to 
recover the result from the compactification of a vacuum solution of
M-theory (eq. \psis\ with $N=7$) we would need to lift the solution \aten\
of ten-dimensional dilaton gravity to a vacuum solution in eleven dimensions.

\subsec{Four-dimensional modular cosmologies from vacuum solutions}

In the following we will be interested in finding exact solutions to the 
field equations derived from the four-dimensional action \ac, that we 
can write in the manifestly $O(N)$-invariant form as
\eqn\oninv{
S=\int d^4 x \sqrt{-g}\left[\CR-2 \partial_{\mu}\Phi^{T}\,\partial^{\mu}
\Phi\right]
}
where we have defined
$$
\Phi \equiv \left(\matrix{\varphi_{1} \cr \vdots \cr \varphi_{N}}\right).
$$
The matter energy-momentum tensor for \oninv\ can be written as a sum of the
corresponding stress-energy tensors for each scalar field\foot{We have normalized the 
energy-momentum tensor in such a way that the Einstein equations are 
$G_{\mu\nu}=T_{\mu\nu}$.}
\eqn\emt{
T_{\mu\nu}=\sum_{i=1}^{N}T^{(i)}_{\mu\nu}=2\left(\partial_{\mu}\Phi^{T}\,
\partial_{\nu} \Phi - \half g_{\mu\nu} \partial_{\sigma}\Phi^{T}\,
\partial^{\sigma}\Phi\right).
}

Let us concentrate our attention on Gowdy-type cosmologies with line
element
\eqn\gowdym{
ds^2= e^{f(t,z)}(-dt^2+dz^2)+K(t,z)\left[ e^{p(t,z)}dx^2+e^{-p(t,z)}dy^2
\right].
}
At first sight, the Gowdy-type coordinates seem an unnecessary complication,
since we will be mostly dealing with cases where both geometries and 
scalar fields are homogeneous. The telling point to use them, nevertheless, is
that: {\it i)} The scalar field equation are linear in the
metric functions. {\it ii)} The evolution of the transversal metric functions
$K(t,z)$ and $p(t,z)$ is decoupled from scalar field dynamics, the 
longitudinal function  $f(t,z)$ being the only one influenced by the 
presence of the scalar fields. And {\it iii)} that due to the presence of a 
six-parameter isometry group $G_6$ which includes the three dimensional 
group $G_3$ acting simply transitively on the three-dimensional surfaces 
of constant curvature in FRW models, the above line element, which has a 
$G_2$ isometry group, naturally includes all three FRW geometries.

Let $f(t,z)=f(t,z)_{\rm vac}$ such that \gowdym\ is a solution of the Einstein
vacuum equations. In this case the functions $p(t,z)$ and $K(t,z)$  satisfy
the following conditions
$$
{\partial\over \partial t}[K(t,z)\dot{p}(t,z)]-
{\partial\over \partial z}[K(t,z)p'(t,z)]  =0 
$$
and 
$$
\ddot{K}(t,z)-K''(t,z)=0.
$$
The idea now is to solve Einstein
equations with the energy-momentum tensor \emt. As it happens with 
a single scalar field (see for example \refs\ralgo), 
the transverse part characterized by the metric functions $K(t,z)$ 
and $p(t,z)$ will be left unchanged. On the other hand, 
the longitudinal function $f(t,z)_{\rm vac}$ is replaced by
$$
f(t,z)_{\rm vac} \longrightarrow f(t,z)_{\rm vac} + f(t,z)_{\rm sc}
$$
and the equations for $f(t,z)_{\rm sc}$ can be written from the Einstein 
equations as
$$
\eqalign{\dot{f}(t,z)_{\rm sc}=& {2K\over K'^{2}-\dot{K}^2}\left(
K'\,T_{tz}-\dot{K} \,T_{tt}\right)\cr
f'(t,z)_{\rm sc} =& {2K\over K'^{2}-\dot{K}^2}\left(
K'\,T_{tt}- \dot{K}\,T_{tz}
\right)
}
$$
where $T_{\mu\nu}$ are the components of the energy-momentum tensor.
Substituting  \emt\ we finally get
\eqn\algor{
\eqalign{\dot{f}(t,z)_{\rm sc}=& {2K\over K'^{2}-\dot{K}^2}\left[
2K'\sum_{i=1}^{N}\dot{\varphi}_{i}\varphi_{i}'
-\dot{K}\left(\sum_{i=1}^{N}\dot{\varphi}_{i}^{2}+
\sum_{i=1}^{N}\varphi_{i}'^{2}\right)\right] \cr
f'(t,z)_{\rm sc} =& {2K\over K'^{2}-\dot{K}^2}\left[
K'\left(\sum_{i=1}^{N}\dot{\varphi}_{i}^{2}+
\sum_{i=1}^{N}\varphi_{i}'^{2}\right)
- 2\dot{K}\sum_{i=1}^{N}\dot{\varphi}_{i}\varphi_{i}'
\right].
}
}
The structure of this expression (sum over each scalar field) 
is the result of the fact that Einstein
equations are linear in the energy-momentum tensor and the 
energy-momentum tensor itself is a sum of the contributions from each 
scalar field.
Expressions \algor\ are invariant under the global 
$O(N)$ symmetry of \ac, as one expects, since a rotation of the
fields by an element of this group does not modify the geometry.
In addition, the scalar fields  $\varphi_{i}$ must satisfy the 
wave equation 
\eqn\wave{
{\partial\over \partial t}[K(t,z)\dot{\varphi}_{i}(t,z)]-
{\partial\over \partial z}[K(t,z)\varphi_{i}'(t,z)] =0,
\hskip 1cm i=1,\ldots,N.
}

Eqs. \algor\ can be used to generalize the algorithm of generation of
string cosmologies given in ref. \refs\rours. Taking  $\varphi_{i}=
\sigma_{i}$ ($i=1,\ldots,6$) and $\varphi_{7}=\phi$, the four
dimensional dilaton, we generate exact solutions
of the Einstein equations with seven scalar fields that, after 
the conformal transformation by $e^{2\phi}$ will provide us with
solutions to the equations derived from the string theory action \st.
After this, we can further transform the resulting four-dimensional
metric by $O(2,2;{\bf R})$ \refs\rours\ or $SL(2,{\bf R})$ \refs\raqmw\
to generate other fields, 
the moduli $\sigma_{i}$ remaining invariant under these transformations.

In addition to this, we can use this generation technique to directly construct
four-dimensional  modular cosmologies representing toroidal compactifications 
of cosmological solutions of M-theory. In the following, we will particularize
our analysis to a certain class of these cosmological solutions  
that render homogeneous cosmologies in four dimensions and use them to 
study how four-dimensional physics can be regularized in eleven dimensions.
This family of solutions will also be useful to study the moduli space
of M-theory on ${\bf T}^{7}$.

\newsec{FRW cosmologies with moduli fields}

We will now apply what we have learnt to the construction of rolling
moduli  cosmologies in four dimensions \refs\rrm. We start with 
a vacuum solution and apply the algorithm
\algor\ with a collection of $N$ time-dependent scalar fields of the 
form
\eqn\sf{
\varphi_{i}(t)=q_{i}\,\varphi_{0}(t), \hskip 1cm i=1,\ldots,N
}
with $q_{i}$ a $N$-tuple of real numbers 
and $\varphi_{0}(t)$ a particular homogeneous
solution to the wave equation \wave. The numbers $\{q_{i}\}$ play 
now the role of coordinates in the moduli space of solutions. From 
\algor\ we get the following equations for
$f(t,z)_{\rm sc}$
\eqn\gen{
\eqalign{\dot{f}(t,z)_{\rm sc}=&-\left(\sum_{i=1}^{N}q_{i}^{2}\right)
 {2K \dot{K}\over K'^{2}-\dot{K}^{2}}\dot{\varphi}_{0}^{2} \cr
f'(t,z)_{\rm sc} =& \left(\sum_{i=1}^{N}q_{i}^{2}\right)
{2KK'\over K'^2-\dot{K}^2}\dot{\varphi}_{0}^{2}.
}
}
Notice that in the family of solutions under consideration, 
the $O(N)$ global symmetry of the Einstein-Klein-Gordon 
action \ac\ acts linearly on the $q_{i}$'s and that the 
numerical prefactor in \gen\ is just the $O(N)$-invariant
metric on the moduli space.

In the generic case, the resulting family of metrics will always
have strong curvature  singularities at some values of the time
coordinate $t$. In what follows we will study what happens to these
singularities from a higher-dimensional point of view, when we 
undo the Kaluza-Klein reduction.

\subsec{Open FRW}
Let us start with the following solution to the vacuum Einstein equations
in four dimensions
\eqn\seedop{
\eqalign{
ds^2_{\rm vac}&=(\sinh{2t})^{-\half}(\cosh{4t}-\cosh{4z})^{3\over 4}
(-dt^2+dz^2)\cr
&+\half\sinh{2t}\sinh{2z}\left(\tanh{z}\,dx^2+
{\rm cotanh\,}z\,dy^2\right)
}
}
and ``dress'' it with the homogeneous scalar fields \sf\ taking
$$
\varphi_{0}(t)= {\sqrt{3}\over 2}\log{\tanh{t}}.
$$
The solution coupled to the $N$ scalar fields is
\eqn\genop{
\eqalign{
ds^2&=(\sinh{2t})^{\half(3\lambda-1)}(\cosh{4t}-
\cosh{4z})^{{3\over 4}(1-\lambda)}(-dt^2+dz^2)\cr 
&+\half\sinh{2t}\sinh{2z}\left(\tanh{z}\,dx^2+
{\rm cotanh\,}z\,dy^2\right),
}
}
where we have defined $\lambda$ as the $O(N)$-invariant quadratic
form
\eqn\conop{
\lambda\equiv \sum_{i=1}^{N}q_{i}^{2}.
}

The interesting feature of these solutions is that 
the dynamics of the scalar fields is only relevant at early times,
saturating to a constant when $t\rightarrow \infty$. One would expect this 
sort of behavior in ``realistic'' models
for the modular fields in string/M-theory cosmology. The scalar fields are
supposed to play an important role in the evolution  
during the early epoch, but are expected to 
be frozen at a certain vacuum expectation value after supersymmetry 
breaking. From that time on, the dynamics of the universe is dominated by 
matter or radiation.

The case with $\lambda=1$ is especially interesting, since we
recover an open FRW universe 
\eqn\frwop{
\eqalign{
ds^2_{\lambda=1}&=\sinh{2t}(-dt^2+dz^2)+
\half\sinh{2t}\sinh{2z}\left(\tanh{z}\,dx^2+
{\rm cotanh\,}z\,dy^2\right) \cr
\varphi_{i}(t)&={\sqrt{3}\over 2}\,q_{i}\log\tanh{t}, \hskip 1cm
{\rm with} \hskip 0.3cm \sum_{i=1}^{N}q_{i}^2=1.
}
}
It is straightforward to check that the metric \frwop\ has
a cosmological singularity at $t=0$ where the curvature scalar
blows up. Actually, the physical properties of the singularity 
 are independent of the particular values $q_{i}$ take, as long 
as the condition \conop\ with $\lambda=1$ is satisfied. This 
is in accordance with the requirement of $O(N)$ global 
invariance of the theory. 

Let us now ``re-construct''  the  $(4+N)$-dimensional theory from 
which \frwop\ is being obtained by dimensional reduction. Here we have
to keep in mind that the scalar fields associated with
the scale factors of the internal torus are related to a linear
combination of the original fields $\varphi_{i}(t)$, as shown 
above. This means that the higher dimensional metric will be
\eqn\nop{
\eqalign{
ds^{2}_{4+N}&=2\,(\sinh{t})^{1-\sum_{i=1}^{N} p_{i}}
(\cosh{t})^{1+\sum_{i=1}^{N} p_{i}}
(-dt^2+dz^2+\sinh^2{z}\,dx^2+\cosh^{2}{z}\,dy^2)\cr
&+\sum_{i=1}^{N}
\tanh^{2p_{i}}{t}\,(dw^{i})^2,
}
}
where we have introduced the new constants $p_{i}=\CD_{ij}q_{j}$, 
with $\CD_{ij}$ the matrix \d. We can now 
rewrite the condition $\lambda=1$ in terms of the new moduli space 
coordinates $\{p_{i}\}$
as
\eqn\metms{
\lambda\equiv \sum_{i=1}^{N}p_{i}^{2}+{2\over 3}\sum_{i<j}^{N}
p_{i}p_{j}=1
}
The new parameters just provide a non-orthogonal system of coordinates
in moduli space. Actually we see from \nop\ that the $p_{i}$'s can
be thought of as a kind of Kasner exponents in the $N$-dimensional 
internal torus when $t\rightarrow 0$. 

The metric \nop\ with $\lambda=1$ is a vacuum solution of Einstein 
equations in $4+N$ dimensions (in fact it is such for any $\lambda$).
Naively, the singularity seems to be 
at  $t=0$ , as it was in the original 
four-dimensional space-time. However, it may happen that this apparent
singularity is just the result of choosing a singular 
coordinate system. To clarify this, we can evaluate the square of the
Riemann curvature tensor $\CR_{abcd}\CR^{abcd}$, which is non-vanishing,
 in the $t\rightarrow 0$ limit
\eqn\curvin{
\CR_{abcd}\CR^{abcd}\sim C(p_{i})t^{2(S-3)}
+\CO[t^{2(S-2)}]
}
where we have written $S=\sum_{i=1}^{N}p_{i}$ and $C(p_{i})$ is defined by 
$$
C(p_{i})={3\over 16}(S-1)^{2}(S^2-2S+5)+\sum_{i=1}^{N}
p_{i}^{2}[3+p_{i}^2+(S-3)(p_{i}+S)]+\sum_{i<j}^{N}p_{i}p_{j}
$$ 
First, let us notice that $S\leq \sqrt{3\lambda}<3$ so the leading
term in \curvin\ will always diverge at $t\rightarrow 0$ limit. 
Thus, in general, we will have curvature singularities in $4+N$ dimensions 
in this limit. However, we find the surprising result that 
whenever $p_{i}=1$, $p_{j}=0$ ($j \neq i$) $C(p_{i})=0$ and actually 
the curvature invariant is regular for all times and given by\foot{Actually,
it is easy to realize that, provided this curvature invariant is finite, 
so are all other scalars constructed from contractions of any number of 
Riemann tensors. In our case this follows from the fact that 
$\CR^{1}_{\,\,a;1b}$ is finite when $t\rightarrow 0$, where the 
index 1 denotes the dynamical internal dimension $w^{1}$.}
\eqn\cir{
\CR_{abcd}\CR^{abcd}={18\over \cosh^{8}{t}}.
}
This corresponds to having only one dynamically non-trivial cycle whose 
scale factor evolve with time. In the four dimensional metric, on the 
other hand, due to the non-diagonal relation between $\{p_{i}\}$
and $\{q_{i}\}$, in general {\it all} scalar fields will have a non-trivial
dynamics. Note, however, that the dimensionally 
reduced model \genop\ is generally 
inhomogeneous for generic $\lambda$, the regular higher-dimensional 
solution falling into a homogeneous and isotropic class.
One might wonder whether higher-dimensional regularity and 
lower-dimensional isotropy are physically linked somehow. 

\subsec{Closed FRW}

We  now pass to discuss  a four-dimensional metric with closed 
spatial sections. The vacuum seed metric is obtained from \seedop\
by replacing hyperbolic functions by their trigonometric counterparts.
We again  couple the space-time to $N$ scalar fields of the form
\sf\ now with 
$$
\varphi_{0}(t)={\sqrt{3}\over 2}\log \tan{t}.
$$
Imposing as before
$\lambda=1$, we are left with the family of four-dimensional
metrics 
\eqn\frwcl{
\eqalign{
ds^2_{\lambda=1}&=\sin{2t}\,(-dt^2+dz^2)+
\half\sin{2t}\,\sin{2z}\,\left(\tan{z}\,dx^2+
{\rm cotan\,}z\,dy^2\right) \cr
\varphi_{i}(t)&={\sqrt{3}\over 2}\,q_{i}\log\tan{t}, \hskip 1cm
{\rm with} \hskip 0.3cm \sum_{i=1}^{N}q_{i}^2=1.
}
}
The coordinate transformation from the standard closed FRW metric to 
\frwcl\ are given in the Appendix of ref. \refs\ralgo, the coordinates 
$z$, $x$ and $y$ used here being just Euler angles for ${\bf S}^{3}$.
In spite of the similarities with the flat case,  the structure
of cosmological singularities is now richer. Studying the 
scalar curvature $\CR$
we find that the metric \frwcl\ has curvature singularities at 
$t=\ell{\pi\over 2}$ with $\ell$ being an integer. We have a bouncing 
universe that evolves from a Big-Bang singularity at $t=\ell\pi$ into a 
Big-Crunch  at $t=(2\ell+1){\pi\over 2}$.

In going to $4+N$ dimensions, we get
\eqn\ncl{
\eqalign{
ds^{2}_{4+N}&=2\,(\sin{t})^{1-\sum_{i=1}^{N} p_{i}}
(\cos{t})^{1+\sum_{i=1}^{N} p_{i}}
(-dt^2+dz^2+\sin^2{z}\,dx^2+\cos^{2}{z}\,dy^2)\cr
&+\sum_{i=1}^{N}
\tan^{2p_{i}}{t}\,(dw^{i})^2,
}
}
where the same definition for the $p_{i}$ as in Sec. 3.1, 
satisfying the condition \metms, has been used. From inspection of \ncl\ we
might be tempted to
infer that the higher dimensional theory will be singular again 
when $t=\ell{\pi\over 2}$. Computing the curvature invariant 
when $t\rightarrow 0^{+}$ we get eq. \curvin, as in the previous
case. However if we now compute the same invariant close to the
Big-Crunch singularity ($t\rightarrow {\pi\over 2}^{-}$) we find
$$
\CR_{abcd}\CR^{abcd}\sim C(-p_{i})t^{-2(S+3)}+\CO[t^{-2(S-2)}].
$$
The situation is somewhat different as compared to the open  case. 
There we found that when the condition $\lambda=1$ is saturated
by a single $p_{i}=1$, the initial Big-Bang singularity disappears 
altogether by going to higher dimensions. Here we find that 
we do not get rid of all curvature singularities, 
but only of ``half" of them. The resulting geometry in this 
case in only singular when $t={\pi \over 2}(2\ell+1)$, whereas 
those at $t=\pi\ell$ are smeared in $4+N$ dimensions. The 
curvature invariant now is given by \cir\ with the hyperbolic cosine
replaced by a trigonometric one. The reverse 
situation happens when we take $p_{i}=-1$ and $p_{j}=0$ ($j\neq 0$):
we avoid singularities located at $t={\pi\over 2}(2\ell+1)$, 
the geometry being the same as for the previous case but now with 
$t\rightarrow t+{\pi\over 2}$.

In ref. \refs\rlw\ a similar situation was noticed for a closed
FRW cosmology that could be resolved into a five-dimensional
black hole interior with just one curvature singularity in the past
or in the future. The difference with the model 
analyzed here is that in the case at hand our higher-dimensional 
universe still has a finite life, although it is doubled with respect
to the one of the four-dimensional geometry.

\subsec{Flat FRW}

Finally, let us briefly analyze the case of cosmological models with
flat spatial sections. The four-dimensional metric and scalar fields
system is given by
\eqn\frwfl{
\eqalign{
ds^2&=2t\,(-dt^2+dz^2+dx^2+dy^2) \cr
\varphi_{i}(t)&={\sqrt{3}\over 2}\,q_{i}\log{t}, \hskip 1cm
{\rm with} \hskip 0.3cm \sum_{i=1}^{N}q_{i}^2=1,
}
}
which is singular when $t=0$. The higher-dimensional version of
this metric according to the  ansatz  \kkan\ is
\eqn\flathd{
ds^2_{4+N}=2t^{1-\sum_{i=1}^{N}p_{i}}(-dt^2+dz^2+dx^2+dz^2)+
\sum_{i=1}^{N} t^{2p_{i}}\,(dw^{i})^{2}
}
We can transform this solution into  a standard Kasner form
by rewriting it in co-moving time coordinates (and re-scaling
spatial coordinates) to give 
\eqn\kas{
ds^{2}_{4+N}=-dT^2+T^{2\alpha_{0}}(dX^2+dY^2+dZ^2)+
\sum_{i=1}^{N} T^{2\alpha_{i}}(dW^{i})^{2},
}
where
\eqn\kasexp{
\alpha_{0}={1-\sum_{j=1}^{N}p_{j}\over 3-\sum_{j=1}^{N}p_{j}}, 
\hskip 1cm \alpha_{i}={2p_{i}\over 3-\sum_{j=1}^{N}p_{j}}
\hskip 0.5cm (i=1,\ldots,N),
}
which can be easily shown to satisfy the Kasner equalities
$$
3\alpha_{0}+\sum_{i=1}^{N}\alpha_{i}=1, \hskip 1cm
3\alpha_{0}^{2}+\sum_{i=1}^{N}\alpha_{i}^{2}=1-{6\,(\lambda-1)\over
\left(3-\sum_{i=1}^{N}p_{i}\right)^2}=1.
$$
Here, we have implemented $\lambda=1$ to get the second condition.

With this expression for the metric we can easily compute the 
curvature invariant, with the result
$$
\CR_{abcd}\CR^{abcd}= {4\over t^{4}}\left[
6\alpha_{0}^{2}(1-\alpha_{0})-3\alpha_{0}^{4}+\sum_{i=1}^{N}\alpha_{i}^{2}
(\alpha_{i}-1)^{2}+\sum_{i<j}^{N}\alpha_{i}^{2}\alpha_{j}^{2}\right]
$$
Using the definition of the Kasner exponents in terms of the 
original $p_{i}$ it is possible to check that the only case in 
which we will have a regular geometry at $t=0$ will again occur if 
$p_{i}=1$ with all other $p_{j}$ ($j\neq i$) vanishing.
In this case the resulting metric is flat. Actually it can be
seen to be $(4+N$)-dimensional 
Minkowski space-time in Rindler coordinates\foot{In ref. \refs\rgh\
an inverse procedure was evoked by dimensionally reducing from a 
five-dimensional flat model to four dimensions to argue that the 
cosmological singularity could be an artifact of dimensional reduction.}.
The non-trivial topology of the internal dimensions will hinder global
identification of the manifold with static  Minkowski solution.

\newsec{M-theory connections}

Let us particularize the study of the cosmological models of the previous 
section to the case with $N=7$ in which they can be interpreted as 
four-dimensional cosmologies arising from a compactification of M-theory 
on ${\bf T}^{7}=({\bf S}^{1})^{7}$ with vanishing three-form. In describing 
the moduli space of cosmological 
solutions coupled to scalar fields of the form \sf\ we have two different 
possibilities. In the four-dimensional theory it seems natural
to take the coordinates $q_{i}$ ($i=1,\ldots,7$) which characterize the 
seven independent scalar fields coupled to gravity. Due to the global 
$O(7)$ symmetry of the low-energy action, there is a natural choice for 
the moduli space metric
$$
I_{4}=\sum_{i=1}^{7}q_{i}^2.
$$
The resulting geometry will only depend on $q_{i}$ through the quadratic 
form $I_{4}$ as seen in equation \gen.

On the other hand, in eleven dimensional language it seems more appropriate 
to change coordinates in moduli space and use instead of $\{q_{i}\}$ the new 
parameters $\{p_{i}\}$ which determine the scale factors of the internal 
compactified dimensions and are related to the original coordinates by 
a non-orthogonal linear transformation, $p_{i}=\CD_{ij}q_{j}$. If we now 
express the moduli space metric in the new coordinates, we find\foot{Since 
we are just performing a change of coordinates, $I_{4}=I_{11}$. We will 
use however different notation to indicate the coordinates used to write 
the quadratic form.}
\eqn\meted{
I_{11}=\sum_{i=1}^{7}p_{i}^{2}+{2\over 3}\sum_{i<j}^{7}p_{i}p_{j}.
}
Incidentally, this moduli space metric is the same as the one obtained in 
\refs\rbfm\ (see also \refs\rrabi) for the compactifications of M-theory 
on a seven-torus with $A_{MNP}=0$ using group theoretical considerations. 
The $O(7)$ global symmetry will act 
linearly on the $p_{i}$'s leaving $I_{11}$ invariant. It is however 
important to notice that the eleven-dimensional geometry do transform under 
$O(7)$, in contrast to the four-dimensional one that was a singlet under 
the action of this group. In eleven dimensions the only transformations 
that leave invariant the space-time metric are permutations of the $p_{i}$'s,
which generates the Weyl subgroup of the mapping class group of the 
seven dimensional torus, $SL(7,{\bf Z})$.

We will consider the family of M-theory metrics obtained by taking $N=7$ 
in \nop. These line elements are vacuum solution of eleven-dimensional 
supergravity 
everywhere in the submanifold of the moduli space defined by $I_{11}=1$. 
We can ask now about the action on this metric of ${\bf G}_{7}\equiv
{\rm Weyl}[E_{7(7)}({\bf Z})]$, 
the subgroup of U-duality transformations preserving the straight torus
${\bf T}^7=({\bf S}^{1})^{7}$ and the vanishing of the three-form 
\refs\rrabi\refs\rbfm, which is generated by  
permutations of the $p_{i}$'s and the so-called 2/5 transformation
\eqn\tf{
(p_{1},\ldots,p_{7})\rightarrow \left(p_{1}-{2s\over 3},p_{2}-{2s\over 3},
p_{3}-{2s\over 3}, p_{3}+{s\over 3},\ldots,p_{7}+{s\over 3}\right)
}
with $s=p_{1}+p_{2}+p_{3}$. It is easy to show that ${\bf G}_{7}$ leaves 
invariant the bilinear form $I_{11}$ and it is thus a discrete finite 
subgroup of $O(7)$ connecting  physically equivalent M-theory vacua. 

In the following we will restrict our analysis to the open FRW metric of 
Sec. 3.1 although most of our results can be extended to the flat case as 
well. 
Acting with the elements of ${\bf G}_{7}$ on the different
solutions characterized by $\{p_{i}\}$ we can look how sensitive M-theory 
physics is
to the semiclassical geometry. The most striking fact we find is that the 
U-duality group 
maps solutions with a Big-bang initial singularity into geometries that 
are regular 
for all values of the cosmic time. For example, the metric
$$
\eqalign{
ds_{11}^{2}&=2\left(\sinh{t}\cosh^{2}{t}\right)^{2\over 3}(-dt^2+dz^2+
\sinh^2{z}\,dx^2+\cosh^2{z}\,dy^2) \cr
&+(\tanh{t})^{2\over 3}\,\sum_{i=1}^{5}(dw^{i})^2
+(\tanh{t})^{-{4\over 3}}\,\sum_{i=6}^{7}(dw^{i})^2
}
$$
which is singular at $t=0$ ($\CR_{abcd}\CR^{abcd}\sim t^{-16/3}$) can be mapped
by ${\bf G}_{7}$ into one of the non-singular M-theory cosmologies using the following
sequence of permutations and 2/5 transformations
$$
\left({1\over 3},{1\over 3},{1\over 3},{1\over 3},{1\over 3},-{2\over 3},-{2\over 3}
\right)\longrightarrow
\left({1\over 3},-{2\over 3},-{2\over 3},{1\over 3},{1\over 3},{1\over 3},
{1\over 3}\right) 
\stackrel{\rm 2/5}{\longrightarrow} (1,0,0,0,0,0,0).
$$
We have concluded that using an U-duality transformation, which is an
exact symmetry of M-theory, we can transform a singular background
into a non-singular one. The obvious bottom line seems to be that, at least in
the low-energy limit, M-theory physics is insensitive to a certain class of 
cosmological singularities.

In the case of the open ($k=-1$) solution, however, the asymptotic 
form of the metrics at large times is to a great 
extend left invariant by the action of ${\bf G}_{7}$. The general 
feature of all solutiones \nop\
when $t\rightarrow \infty$ is that the open three dimensional space-time 
inflates while the internal torus reach a constant volume, namely
$$
{{\rm Vol}_{\rm 3D}\over {\rm Vol}_{{\bf T}^{7}}}\sim \tau^{3},
$$
with $\tau$ the co-moving time. Physically, what we have is a cosmological 
solution that
evolves from an eleven-dimensional regime in which all scale factors are 
of the same size 
into an asymptotic state with describes a ``large'' four-dimensional 
expanding universe and
a ``small'' static seven-dimensional torus (cf. \refs\rdc\refs\rdem).

In the case of the flat solution, on the other hand, the large-time
behavior of the metric is not universal. As we saw in Sec. 3.3, in 
this case we can rewrite the metric in the Kasner form \kas\ 
with exponents given by \kasexp. The U-duality group ${\bf G}_{7}$ 
acts on these exponents in a rather complicated way through the 
transformation of the moduli space coordinates $\{p_{i}\}$ (cf. \refs\rbfm). 
Since the sum $\sum_{i=1}^{7}p_{i}$ is not invariant under the 2/5 transformation, 
the asymptotic form of the metrics \flathd\ or \kas\ will be sensitive to rotations by
elements of the U-duality group. In addition to this, the internal manifold
will not remain ``small'' in general. We will have the usual behavior
of any Kasner universe, with at least one contracting direction and
a number of expanding ones.

The main characteristic of the non-singular M-theory cosmologies 
that we have obtained is that in all cases there is only one time-varying 
coordinate, while the remaining ones are constant during the evolution 
of the Universe. In trying to make a string theory interpretation of such 
solutions it seems natural to identify the dynamical coordinate with 
the eleventh dimension associated with the dilaton field. If we do
so and perform a dimensional reduction of the open ($k=-1$) 
regular eleven-dimensional 
solution down to ten dimensions, the resulting metric and dilaton 
in string frame are\foot{The appropriate Kaluza-Klein ansatz 
in this case is $ds^2_{11}=
e^{-{2\over3}\phi}ds^2_{10}+e^{{4\over 3}\phi}dw^2$.}
$$
\eqalign{
ds^{2}_{\rm string}&=\sinh{2t}\,(-dt^2+dz^2+\sinh^2{z}\,dx^2+
\cosh^{2}z\, dy^2)+
\tanh{t}\sum_{i=1}^{6} (dw^{i})^{2}, \cr
\phi(t)&= {3\over 2}\log\tanh{t}
}
$$
which corresponds to a string background with three open spatial dimensions and
another six compactified on a torus whose dynamics is characterized
by a global breathing mode (a situation extensively studied in the 
Kaluza-Klein and string 
cosmology literature \refs\rkkbm\refs\rbbang\refs\rrm). 
The geometry is singular when $t\rightarrow 0$ but, on
the other, at large times we find again an inflating three-dimensional space 
together with a frozen internal torus, now also with a constant dilaton.
Notice, however, that M-theory U-duality maps these kind of singular 
string cosmologies
into the regular ones that are obtained by identifying the dilaton field 
with one of the
static circles. These transformations involve the interchange of the 
dilaton with
other moduli fields and thus are intrinsically
non-perturbative from the point of 
view of string theory. 

Incidentally, let us remark that because of the invariance of $I_{11}\equiv 
\lambda$ under $O(7)$, the family of solutions \genop\ is stable under
the U-duality group which thus maps solutions into solutions (cf. \refs\rbfm). 
This is what
one should expect, since ${\bf G}_{7}$ is a symmetry of M-theory on a straight 
torus with a vanishing three-form.  

\newsec{Concluding remarks}

In this paper we have been concerned with four-dimensional cosmologies 
that arise from the compactification of string/M-theory on a straight torus. 
We have given a general algorithm to generate these modular cosmologies in 
four dimensions with two commuting isometries. Using this technique 
we have also constructed FRW metrics and observed that, although they are
singular in four dimensions, in the open ($k=-1$) and flat ($k=0$) cases 
their lifted higher-dimensional vacuum images have a regular curvature 
invariant $\CR_{abcd}\CR^{abcd}$. In the case of closed FRW cosmologies
($k=1$), singularities are not
removed but the lifetime of the higher-dimensional universe (i.e. the 
time elapsed from the Big-Bang to the Big-Crunch) is twice that of their 
four-dimensional versions.

It is important to stress, however, that in general the homogeneous character
of the scale factors  of the extra dimensions derived from \sf\ does not 
ensure the homogeneity of 
the reduced four-dimensional cosmology. Given the absence of a singularity-free
higher-dimensional ``parent'' cosmology for the 
four-dimensional inhomogeneous solutions,
it is remarkable that the regularity condition for the space-time 
in higher dimensions translates itself into isotropy of the four-dimensional 
solution. Thus, the initially 
isotropic universe could be naturally chosen by the requirement of regularity 
of the higher-dimensional space-time. 

We have also studied four-dimensional modular cosmologies related to toroidal
compactifications of string/M-theory. In the case of M-theory on a straight
seven-torus with vanishing three form we have found open and flat singular 
metrics that can be rotated into regular ones by the elements of  
the U-duality group
${\bf G}_{7}$, thus indicating that the physics of a certain class of 
singular universe can be described in terms of a dual regular geometry. 
In the case of homogeneous open universes, however, the large time 
asymptotic behavior
of the metric is insensitive to the action of ${\bf G}_{7}$. In all cases
we find an inflating three-dimensional space with $k=-1$ and an internal torus 
whose size asymptotically reaches a constant value. It is remarkable that 
U-duality is able to relate geometries with such a different behavior
near $t=0$ without modifying the dynamics of the universe at large times.
Incidentally, this stabilization of the internal dimensions is produced 
by the presence of the positive spatial curvature and it is absent in 
the case of the flat solutions.

Because of the $O(7)$ invariance of the dimensionally reduced cosmologies, all 
the eleven-dimensional metrics labeled by $\{p_{i}\}$ and the same value of 
$I_{11}$ will produce exactly
the same geometry in four-dimensions, although it will be coupled to scalar 
fields
with different amplitudes in each case. In particular, using the classical 
symmetry of the four dimensional action {\it any} solution in our family 
with $I_{4}=1$ can be rotated 
into one with a regular higher-dimensional ``parent'' metric. However, 
since M-theory in 
this background is not invariant under the full group $O(7)$, one expects 
that quantum
effects will break this global symmetry down to the U-duality group
${\bf G}_{7}$ which preserves the lattice of charges. The situation is 
completely analogous to the classical invariance
of the four-dimensional effective action under $O(6,6;{\bf R})$ \refs\rms, 
a symmetry that is broken down to the T-duality group $O(6,6;{\bf Z})$ by 
string effects.

Finally, it would be interesting to check whether our results can be extended
to compactifications of M-theory on a generic seven-torus with a non-vanishing
three form. From our analysis it is clear at least that there will be 
solutions of this kind that can be transformed into non-singular universes 
by U-duality, as can be seen by acting on the regular metric with the Borel 
generators of $E_{7(7)}(\bf Z)$ \refs\rop.

\newsec{Acknowledgements}

We thank J.L.F. Barb\'on and J.L. Ma\~nes for interesting discussions. 
M.A.V.-M. wishes also
to thank K. Skenderis and E. Verlinde for discussions on ref. \refs\rbfm.
The work of A.F. has been supported by 
Spanish Science Ministry Grant 172.310-0250/96 and the University of the
Basque Country Grant UPV 172.310-EB150/98 and 
that of M.A.V.-M. by FOM ({\it Fundamenteel Ordenzoek der Materie}) and 
by the University of the Basque Country Grant UPV 063.310-EB187/98.

\listrefs

\bye